# Dual-Band Meta-Absorber for Reduced Cross Talk between 5G MIMO Antenna Arrays

Amjad Nadeem
Electrical Engineering Department, Al Jouf university, Saudi Arabia

*Abstract*— Metamaterials have great potential to construct ultrathin, miniaturized, and cost-effective microwave and optical devices for future wireless communication systems. Here, a dual-band meta-absorber comprising four arrows-shaped unit cells is investigated. The absorption features are numerically studied and analyzed in a microwave frequency spectrum from 28 - 36 GHz. The proposed meta-absorber illustrates the two independent absorption peaks of 96% and 99.99% at the corresponding operating points of 31.11 GHz and 33.77 GHz, respectively, within the 5G spectrum. Due to structure symmetry, the proposed meta-absorber exhibits insensitive polarization performance under the influence of different polarization angles of the incoming EM waves. The absorption characteristics of the absorber are also analyzed for different obliquity of the incident EM waves, and their absorption peaks are tuned under these operating conditions. Further, to realize the absorption behavior of this absorbing device, the surface electric field profile and current distributions are also studied at the aforementioned two operating frequency points. This absorber type remains prudent to minimize the coupling and cross-talk between the elements of the MIMO antenna operating in the 5G band.

## I. Introduction

Metamaterials are in the limelight as they offer exotic phenomena like backward waves, negative refraction, Cherenkov radiations and reverse Doppler Effect. Based on these properties, different R&D communities are exploring their potential applications in lensing, holography, cloaking, ultrasensitive bio-sensing and so forth. Meta-absorber is another application of this series where subwavelength scale unit cells have the capability to absorb the incident light while exhibiting zero reflection. The resonance frequency impedance of metasurface matches with free space impedance, leading to maximum absorption and almost zero reflection. Such meta-absorbers have widespread applications in wireless-power-transfer (VPT), radar cross-section reduction (RCSR), thermo-photovoltaic, stealth, and others.

Design Matrix for modern antennas relay on wide bandwidth, high throughput, low cost, and miniaturization of structure. Over the past few years, multi-input-multi-output (MIMO) antennas have gained widespread interest in meeting the design matrix requirements. In MIMO configuration, several antenna arrays are usually multiplexed so that each antenna can be connected with a separate input port for excitation [1, 2].

Here in this paper, we have proposed a novel four arrows shaped PCB printed meta-absorber. The proposed design exhibits two independent absorption peaks of 96% and 99.99% in the 5G band. The proposed unit cell is a three-layer structure with copper-made symmetric arrows embedded over an FR4 substrate backed by a thin copper layer, which behaves as a perfect reflector. Electric field intensities and surface current distributions are also explored to study the absorption mechanism. To the best of our knowledge, symmetric four arrows shaped unit cell is a novel design and was firstly reported in the present communication. Due to perfect absorption, the proposed system has potential applications in multi-band sensing, filtering, and 5G communication.

## II. Design Technique

Fig. 1 (a) & (b) illustrates the schematic of four arrows shaped meta-absorber. The designed meta-absorber is composed of the sandwich structure. The top layer is made of four arrows shaped structures pointing toward each other. Lossy dielectric material FR4 is used in the middle as a substrate which is backed by a thick metal layer. Copper annealed is used for making the top and bottom layer with thickness $t_m = t_g = 0.035$ mm. Dimensions of other physical parameters are $P = 5$ mm, $L = 4.4$ mm, $w = 0.5$ mm, $t_d = 1.6$ mm and $g = 0.065$ mm.

All the designing and simulations are done using the commercially available simulation tool CST Microwave studio. To be more explicit, a built-in frequency-domain solver is applied to extract the s-parameters of the designed meta-absorber. In the simulation setup, unit cell boundary conditions are employed in *x-y*-directions, while open add space in *z*-direction along the incident wave traveling. When an electromagnetic wave shines on the meta-absorber, it penetrates inside due to impedance matching between the top layer of the meta-absorber and free space. This incoming wave gets trapped in the dielectric layer. The top and bottom layers help in wave confinement within the structure.

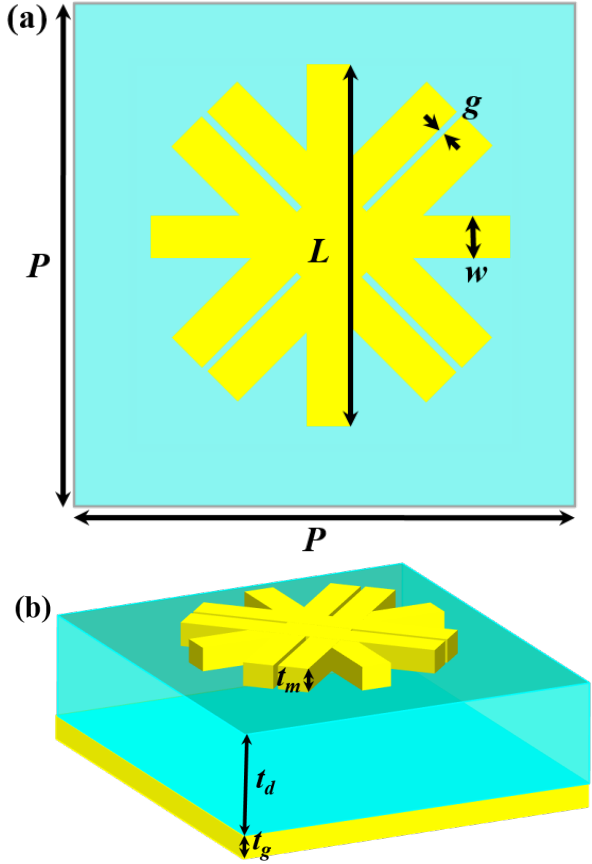

Fig. 1. Schematic of the proposed unit cell. (a) Two-dimensional front view. (b) Three-dimensional view.

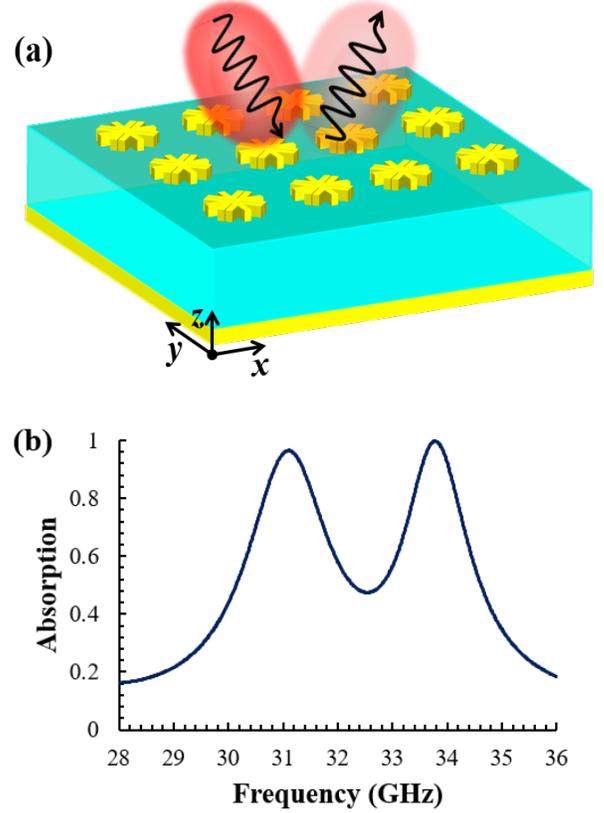

Fig. 2. Schematic and absorption of proposed meta-absorber. (a) Schematic of proposed meta-absorber. (b) Simulated absorption spectra under the normal incidence of electromagnetic wave.

## III. RESULT AND DISCUSSION

Under normal incidence of electromagnetic waves, the absorption phenomenon can be abstracted by utilizing the material's permittivity ($\varepsilon$) and permeability ($\mu$). Impedance matching is achieved when these aforementioned parameters are equalized. At resonance frequencies, the impedance matching condition is achieved by properly optimizing the physical dimensions of the unit cell. So at resonance $Z_m = \sqrt{\mu/\varepsilon} \approx Z_o$, where $Z_o$ and $Z_m$ are impedance parameters of free space and meta-absorber respectively. Therefore, the reflection coefficient vanishes, i.e., $\Gamma = ((Z_m - Z_o)/(Z_m + Z_o)) \approx 0$ as the bottom layer is metallic, so the transmission coefficient (T) also becomes zero. Resultantly, absorption can be defined as $A = 1 - |\Gamma|$.

Fig. 2 depicts the schematic of the meta-absorber along with the absorption results under normal incidence. Absorption analyses are done for the 28 - 36 GHz frequency band. The proposed meta-absorber manifests two independent absorption peaks of 96% and 99.99% at the frequencies of 31.11 GHz and 33.77 GHz, respectively, within the 5G spectrum.

Angular dependence of the absorption spectrum for the proposed meta-absorber is also studied. An electromagnetic wave impinged from different oblique incidences with fixed step width. Obtained frequency-dependent results are displayed in Fig. 3 (a). Maximum absorption is achieved under normal incidence. There is no significant drop in absorption for oblique incidences (10° - 40°), but their resonance peaks shift slightly. However, for θ > 40° incidences, absorption drastically reduces. Absorption spectra under different polarization angles ($\varphi$) are also investigated. The performance of the proposed meta-absorber for $\varphi$ = 0° - 90° with a step-width of 30° is depicted in Fig. 3 (b). Due to the four-fold symmetry of the unit cell, obtained results are true overlaps with each other. Therefore, the proposed meta-absorber is polarization insensitive.

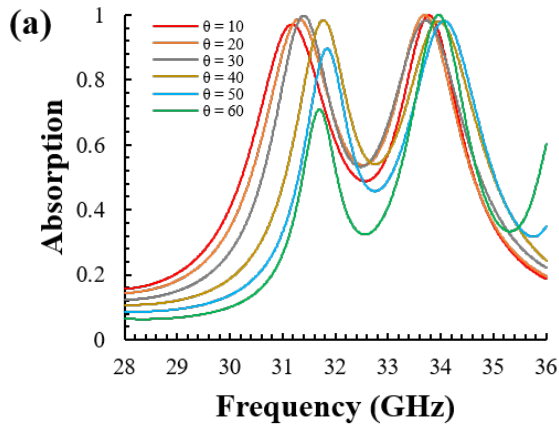

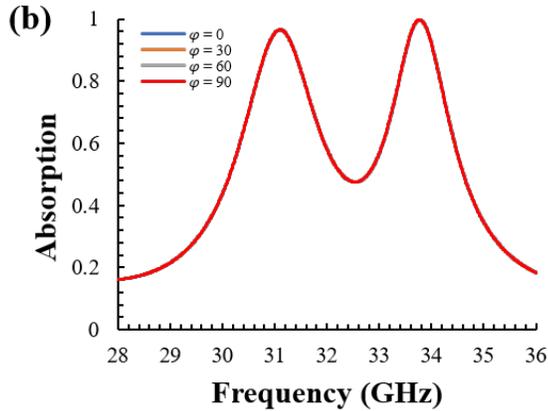

Fig. 3. Frequency-dependent absorption curves for proposed meta-absorber. (a) At different incidence angles. (b) At different values of polarization angles $\varphi$.

Electric (E-) field distribution and current density are studied at resonant frequencies for normal incidence to understand the phenomenon behind absorption. The top surface E-field distribution at 31.11 GHz and 33.77 GHz is depicted in Fig. 4 (a) and (b), respectively. In Fig. 4 (a), it's clear that most of the E-field is accumulated on the edges of the middle bar of two vertically aligned arrows. This determines that this vertically aligned rib is mainly responsible for first resonance. Fig. 4 (b) illustrates that most of the E-field is concentrated on the edges of vertically aligned arrows. Therefore, the central and adjacent ribs' accumulated response is responsible for the second resonance peak.

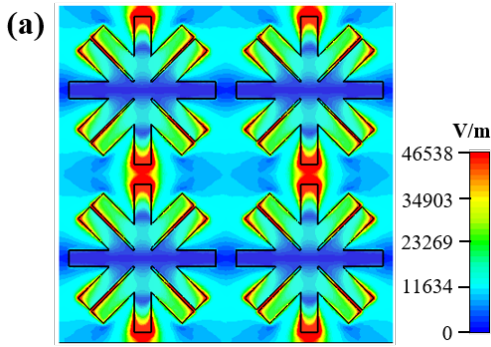

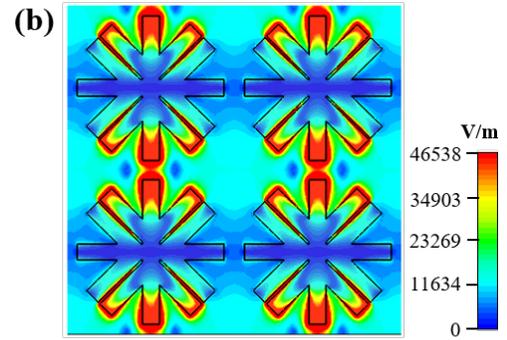

Fig. 4. Electric field distribution for proposed meta-absorber. (a) At 31.11 GHz and (b) At 33.77 GHz.

## IV. CONCLUSION

In the aforementioned discussion, four arrows shaped dual-band meta-absorber is designed and simulated, producing 96% and 99% absorption peaks in the 5G band. Due to the four-fold symmetry of the unit cell proposed meta-absorber is polarization insensitive. It produces stable absorption peaks for oblique incidence angles. E-field and surface current distributions are studied to deeply understand absorption mechanisms. Due to perfect absorption, the proposed meta-absorber's insensitivity of polarization and incident angle is an ideal candidate for numerous 5G applications like MIMO antenna isolation, attenuation, and filtering.